\def\kms{km~s$^{-1}$}
\def\etal{{\it et al.}}

\def\sqd{{deg$^{2}$}}

\documentclass[
    ,final            
  ]
  {aipproc}

\layoutstyle{6x9}

\newcommand\doingARLO[2][]{%
  \ifx\mmref\undefined #1\else #2\fi
}

\begin{document}

\title 
      {HI Cosmology at $z=0$: a Brief Review}


\author{Riccardo Giovanelli}{
  address={Space Sciences Bldg., Cornell University, Ithaca, NY 14853},
  email={riccardo@asrto.cornell.edu}
}

\copyrightyear  {2008}

\begin{abstract}
Extragalactic HI astronomy is half century old. Its maturity dramatically
increased in the 1970s, with the commissioning of new powerful facilities.
Its contributions to Cosmology are important, from the observation of 
galaxy rotation curves that showed the presence of dark matter in galaxies,
to the measurement of cosmological parameters and the mapping of the large
scale structure of the Universe. The Arecibo telecope has played a key role
in these developments. It is also currently engaged in a number of experiments
that utilize its L--band feed array to map thousands of square degrees
of the sky and obtain the most sensitive large--scale view of the low $z$ HI
Universe. 
\end{abstract}

\date{\today}

\maketitle

\section{Introduction}

I have been asked by the organizers of this workshop to devote opening
remarks to a brief review of achievements in extragalactic HI astronomy 
and how these set the stage for current efforts in the field, particularly
the ALFALFA survey. Because of the judgement 
passed not long ago by a national review committee, a dark shadow has been cast 
on the future --- and very existence in the next decade -- of the most beautiful 
of radio observatories, which hosts this workshop today. Finding myself in 
disagreement with some of the conclusions of said review, I make no apology for 
the fact that my report will lean towards the body of discovery Arecibo has 
contributed in extragalactic spectroscopy, while in the last session of this
workshop, I chance some predictions on the possible large--scale projects
Arecibo may be able to attack in the next decade.

Baryons make up about 4.5\% of the mass/energy budget of the Universe, and only
1/6 of its matter density. At $z=0$ the vast majority of baryons are thought to
exist in the form of coronal and intergalactic gas, at temperatures $>10^{5}$ K; 
$\Omega_{stars}$ is a tiny 0.0027 and $\Omega_{cold~gas}$ an even smaller 0.0008, 
of which a bit over half is neutral Hydrogen~\cite{omega}, the target of
this workshop's observational concerns. 
This unimpressive budgetary datum could well prompt the question: why do we care
about extragalactic radio spectroscopy? 
First, HI is easy to detect at 21 cm wavelength, most of the emission originates in
optically thin regions and cold gas masses are reliably measured; the abundance of
cold gas is a reliable indicator of star forming potential for an extragalactic
system. Second, the distribution of HI, which can be found at larger galactocentric
distances than other easily
detectable components in a galaxy, makes it an excellent tracer of the large--scale
dynamics of its host. Third, scaling relations of disks, such as that between luminosity
and rotational width, make HI measurements good cosmological tools: for example in
the measurement of $H_\circ$, peculiar velocities, the convergence depth of the Universe
and the local matter density field. Fourth, because of its presence at relatively
large galactocentric distances, HI is more vulnerable than the stellar component
to external influences and thus
constitutes a good tracer of tidal interactions, mergers and other environmental
effects. Fifth, it can be the dominant baryonic component in low mass galaxies and
thus provide a reliable census of low mass systems in the galactic hierarchy.

The 21 cm line of neutral Hydrogen was first detected in 1951 from Milky Way
clouds and in 1953 from the Magellanic Clouds. It took nearly two decades of
pioneering work, largely by M.S. Roberts, to bring the number of galaxies
detected in HI to exceed one hundred \cite{roberts75}. During this period,
the near totality of extragalactic HI observations were made with single
dishes, at Green Bank, Nan\c cay, Effelsberg and Owens Valley. In the 1970s,
mature, new aperture synthesis telescopes at Westerbork and the VLA were
commissioned, followed two decades later by the GMRT, the largest synthesis 
telescope currently operating at 21cm. In the early 1970s, the primary reflecting surface 
of the Arecibo telescope was upgraded, making operation at L--band possible.
At the same time, lower $T_{sys}$ receivers and broad band spectrometers 
also became available.
The sensitivity of the 305m telescope provided the complementary counterpart
to the high angular resolution capabilities of the synthesis intruments.
Sensitivity was unfortunately not matched initially at Arecibo by breadth of 
field of view. With a field of view of a single--beam of $\sim 3.3'$, 
unbiased, sensitive, blind, large--scale surveys remained impractical.  
This situation changed with the
implementation of focal plane arrays in large single dishes. The first
effective use of that technology for HI spectroscopy took place in the 1990s
at Parkes and it yielded the important HIPASS catalog \cite{hipass} ~of about 
5000 HI sources. ALFA, a seven--feed focal plane array, was commissioned at 
Arecibo in 2004, making possible full--fledged, wide solid angle survey 
operations, which started in early 2005.

\section{Discovery Highlights}

In 1973, in reporting the observations of apparently flat rotation curves of three
nearby galaxies, Roberts \& Rots\cite{rr} wrote: {\it ``The shapes of the rotation
curves at large radii indicate a significant amount of matter at these large
distances and imply that spiral galaxies are larger than found from [optical]
photometric measurements''}. Through HI observations they had discovered dark matter 
in galaxies. Since then, thanks to the fact that HI is detectable to larger 
galactocentric distances than other baryonic forms, HI measurements have been the 
technique of choice for tracing the mass distribution at large galactocentric radii.
This field found fertile grounds at the WSRT, Arecibo and the VLA. The first ``outing''
in high $z$ territory of the HI line was its detection in absorption in 3C286
at $z=0.692$ by Brown \& Roberts\cite{br73}.

In 1972, Gunn \& Gott\cite{gg} proposed that ram pressure would have an important
effect on the evolution og galaxies in clusters. After pioneering work at Jodrell
by Davies \& Lewis\cite{dandl73}, the effect was detected at Nan\c cay\cite{cha}.
The correlation between truncation of the gas disks with HI deficiency was
revealed at Arecibo(\cite{gio3},\cite{hay2}), and the wealth of Arecibo data
lead to a clear match between the distributions of the hot IGM and HI deficiency
\cite{sol}. On a complementary path, synthesis observations\cite{cay} illustrated 
the details of the galaxy--IGM interaction, more recently surveyed at the VLA
by Kenney, van Gorkom and collaborators (VIVA) and as described by Jacqueline
in these  proceedings.

In the 1980s, Arecibo became a productive redshift machine. As evidence for
structure in the distribution of galaxies on scales well in excess that of 
clusters started to mount in the late 1970s, a program was started to map putative
supercluster structures by measuring the redshifts of large, optically selected
samples of galaxies. The data gathered by Giovanelli, Haynes and collaborators 
through that effort, including nearly 10,000 redshifts and HI parameters, are 
accessible in digital form through an archive maintained at Cornell University
\footnote{http://arecibo.tc.cornell.edu/hiarchive/}. These data provided a clear
view of the filamentary character of the large--scale structure of the Universe,
especially in the characterization of the Pisces--Perseus supercluster.

The relationship between optical luminosity and rotational velocity of spiral
disks, first proposed by Tully \& Fisher\cite{tf} to be usable as a tool to
estimate extragalactic distances independent on redshift, became widely used
in the 1980s and 1990s, both for the measurement of the Hubble constant and
of bbpeculiar velocities. Galaxies' peculiar velocities arise from 
gravitational perturbations due to inhomogeneities in the density field. The
peculiar velocity at a given location in space is the cumulative effect of
such perturbations, out to a maximum distance often referred to as the
``convergence depth''. Pertubations originating at locations farther than the
convergence depth have negligible amplitude, as they colllectively balance
against each other. The determination of the convergence depth, which can be
measured by gauging the reflex motion of the Local Group with respect to 
galaxies populating spherical shells of progressively increasing radius, is 
a proxy for that of the scale of a fair sample of the Universe.
The largest sample of spiral galaxy peculiar velocities is known as SFI++,
consisting of a data base of I--band photometry and 21cm line and H$\alpha$
spectroscopy of approximately 4500 galaxies\cite{hay99}. About half of the 
SFI++ sample lies within the Arecibo telescope Declination range. A complementary
sample of cluster galaxies allowed the accurate determination of a template
Luminosity--linewidth relation\cite{sci}, which has been used to map the
peculiar velocity field, determine the convergence depth and confirm the Doppler
nature of the CMB dipole\cite{cdepth}, measure the Hubble
constant\cite{hnot} ~and $\Omega_{mass}$\cite{borgani97}.

In the mid--to--late 1990s, the Arecibo telescope's astronomical activities
were largely suspended as most of its line feeds --- early devices for correction 
of the aberration resulting from its spherical primary ---, were replaced with a 
Gregorian subreflector system. In the period during which partial operation was 
possible, two pioneering surveys were carried out in the 21 cm line: AHISS by 
Briggs and co--workers\cite{zwaan} and ADBS\cite{ross}. Although they covered 
relatively small solid angles of sky --- respectively 13 and 430 square
degrees for AHISS and ADBS, they were the first in surveying unbiasedly
the extragalactic HI Universe with interesting sensitivities. They
detected respectively 65 and 265 HI sources, or 5 and 0.6 sources per
square degree. The two surveys yielded very different HI mass functions,
especially at the faint end. These differences arise mainly from the
small numbers of sampled objects and the uncertainties in the distances
of the faintest objects in each survey: for those, the (unknown) 
individual peculiar velocities can be comparable with or even larger than
the Hubble velocity. HI masses can thus be uncertain by factors of several.

\begin{figure}[]
\resizebox{0.7\textwidth}{!}{\includegraphics{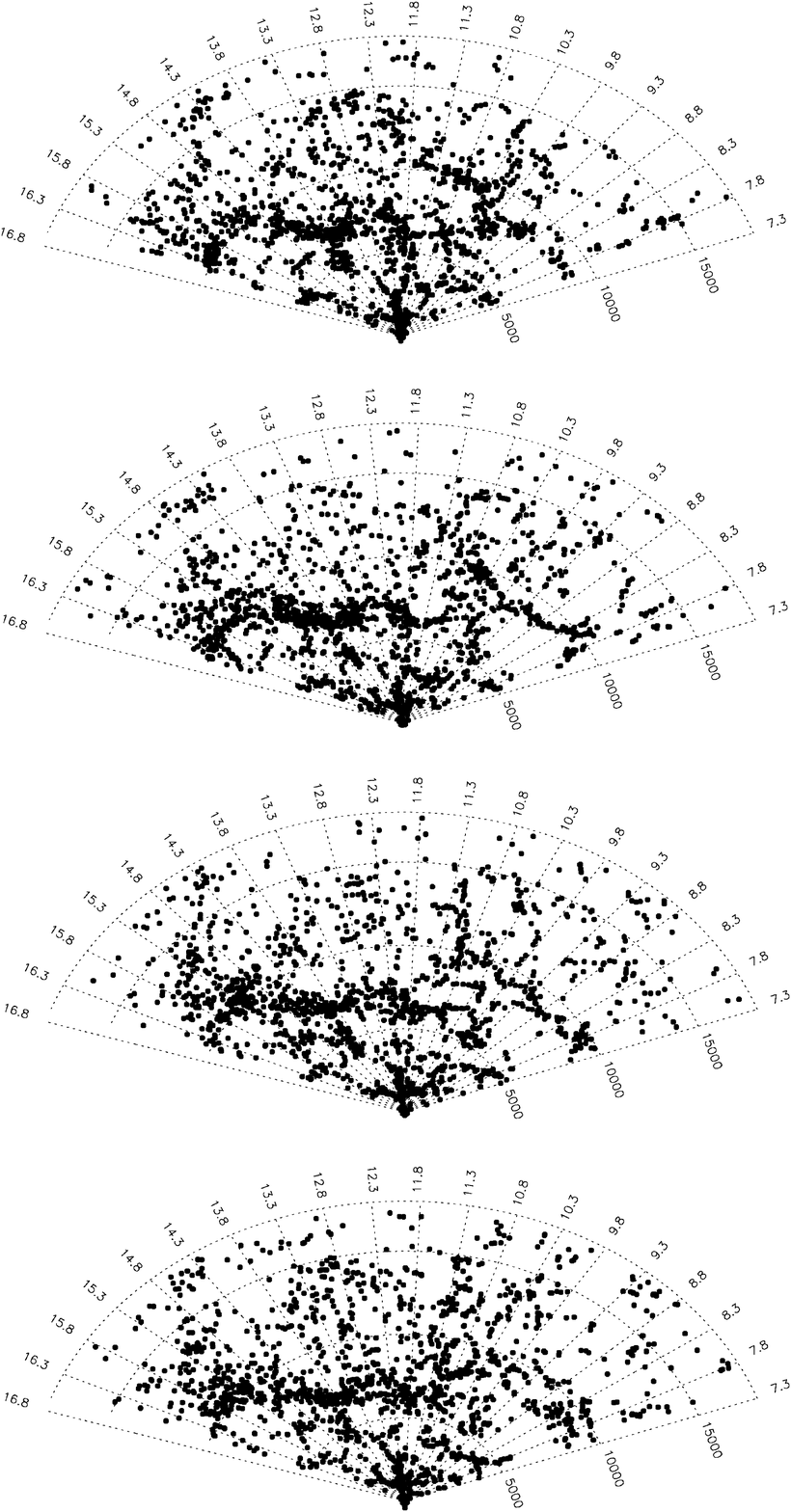}}
\caption{
R.A.[hours] vs $cz$[\kms] wedge diagrams of four independent data sets,
corresponding to declination ranges, top to bottom: [13$^\circ$--16$^\circ$],
[10$^\circ$--13$^\circ$],[07$^\circ$--10$^\circ$],[04$^\circ$--07$^\circ$];
the R.A. range is the same in all graphs, between 07:30 and 16:30 hours.
ALFALFA detections are plotted. Each diagram includes approx. 2000
objects. The sky region corresponds to approximately 22\% of the projected
survey solid angle of ALFALFA. Note that due to RFI, ALFALFA is effectively 
blind in the redshift range between approximately 15000 and 16000 \kms.
}
\label{4cones}
\end{figure}

While Arecibo was being upgraded, a feed array was being commissioned
at Parkes, which eventually delivered the widest extragalactic
HI survey to date, HIPASS\cite{hipass}. Now complete, HIPASS has surveyed
30,000 square degrees of sky, including the full southern hemisphere.
With an average sensitivity of 13 mJy per beam at 18 \kms ~spectral
resolution, it produced a catalog of approximately 5000 HI sources, a
sky density of 0.17 sources per square degree. Table \ref{tab1} shows
a comparison of various blind HI surveys, past and current.

\begin{table}[!t]
\caption{Comparison of Blind HI Surveys}
\begin{tabular}{lccccccc}
\hline
  \tablehead{1}{r}{i}{Survey}
  & \tablehead{1}{r}{b}{Beam(')}
  & \tablehead{1}{r}{b}{Solid angle (\sqd)}
  & \tablehead{1}{r}{b}{res(\kms)}
  & \tablehead{1}{r}{b}{\tablenote*{rms in mJy per beam uniformly referred at 18 \kms ~resolution}}
  & \tablehead{1}{r}{b}{V$_{med}$(\kms)}
  & \tablehead{1}{r}{b}{N$_{det}$}
  & \tablehead{1}{r}{b}{Ref}
\\
\hline
AHISS   & 3.3 &    13 & 16 & 0.7 & 4800 &   65 &  \cite{zwaan} \\
ADBS    & 3.3 &   430 & 34 & 3.3 & 3300 &  265 &  \cite{ross} \\
WSRT    & 49. &  1800 & 17 & 18  & 4000 &  155 &  \cite{braun} \\
HIPASS  & 15. & 30000 & 18 & 13  & 2800 & 5000 &  \cite{hipass} \\
HI-ZOA  & 15. &  1840 & 18 & 13  & 2800 &  110 &  \cite{zoa} \\
HIDEEP  & 15. &    32 & 18 & 3.2 & 5000 &  129 &  \cite{hideep}\\
HIJASS  & 12. &  1115 & 18 & 13  & \tablenote*{HIJASS has a gap in velocity coverage between 4500-7500
\kms, caused by RFI}                    &  222 &  \cite{lang} \\
J-Virgo & 12. &    32 & 18 &  4  & 1900 &   31 &  \cite{jvirgo} \\
AGES    & 3.5 &   200 & 11 & 0.7 & 12000&      &  \cite{ages} \\
ALFALFA & 3.5 &  7074 & 11\tablenote*{raw data at 5.5 \kms} 
                           & 1.7 & 7800 &$>$25000& \cite{a1} \\
\hline
\end{tabular}
\label{tab1}
\end{table}

\begin{figure}[]
\resizebox{1.1\textwidth}{!}{\includegraphics{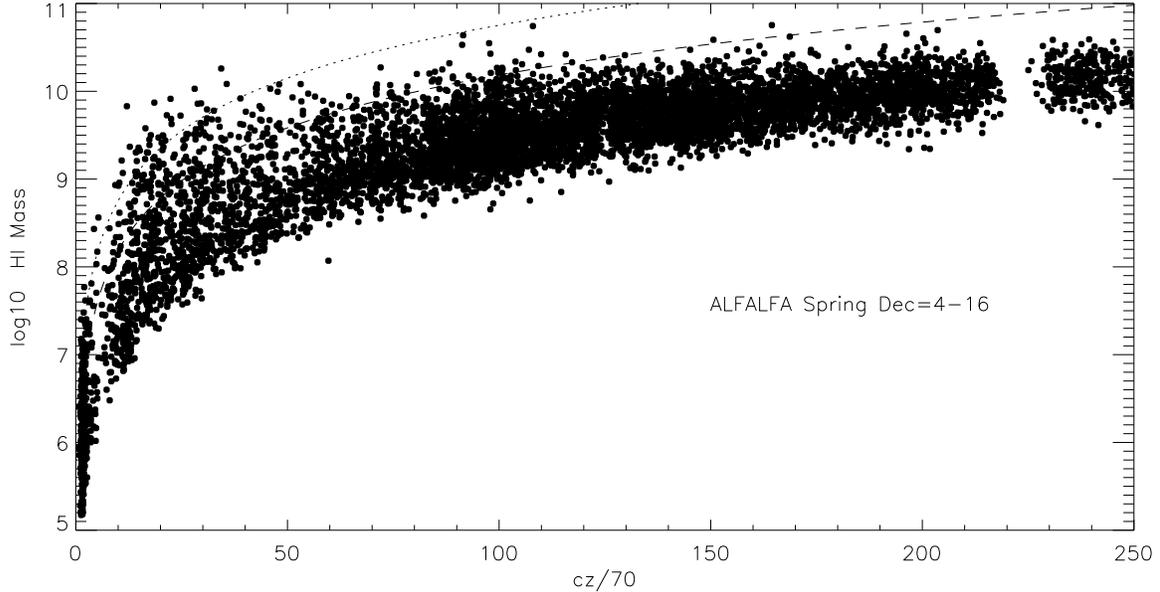}}
\caption{
Sp\" anhauer plot of 7938 HI sources in the region R.A.=[7.5$^h$--16.5$^h$],
Dec=[4$^\circ$--16$^\circ$]. Nearby objects of uncertain distance, possibly
galactic HVCs, ``pollute'' the plot at $M_{HI}<10^{6.5}$.
The two smooth lines identify the completeness limit for sources of 
200 \kms ~linewidth (dotted) and the overall detection limit (dashed) 
for the HIPASS survey. Note that due to RFI, ALFALFA is effectively 
blind in the redshift range between approximately 15000 and 16000 \kms.
}
\label{span}
\end{figure}

\section{ALFALFA}

The most ambitious among currently ongoing, blind HI extragalactic surveys
is ALFALFA (the {\it Arecibo Legacy Fast ALFA} survey)\cite{a1}, which will cover
7000 square degrees of sky and is on course to detect more than 25,000 HI 
sources, over a 100 MHz bandwidth with 25 kHz ($\sim 5.5$ \kms) spectral resolution.

Figure \ref{4cones} shows the redshift distribution of a 22\% section of the 
ALFALFA survey. The median $cz$ is near 8000 \kms, the typical scalelength
of baryonic acoustic oscillations. ALFALFA is the only large--scale HI survey
that samples a fair volume of the Universe (HIPASS' median $cz$ is less than
3000 \kms). A comparison of ALFALFA and HIPASS is clear from inspection of 
Figure \ref{span}: of the $\sim 8000$ ALFALFA sources displayed, HIPASS would detect
fewer than a few dozens to its completeness level, and a few percent to its overall
detection limit. Moreover, a large fraction of HIPASS sources suffer from
confusion because of the large Parkes telescope beam, making the identification 
of optical counterparts difficult and
often impossible without follow--up, higher resolution HI observations. The
smaller Arecibo beam largely obviates the problem: more than 95\% of ALFALFA
sources can be unambiguously associated with the correct optical counterpart.
HIPASS detected fewer than two dozen sources with HI mass $<10^{7.3}$ solar;
ALFALFA is on track to detecting a few hundred.

ALFALFA will allow an accurate determination of the HI mass function to
unprecedented low levels; elucidating the large--scale structure characteristics 
of HI sources, their impact on the ``void problem''; mapping the local mass 
density field; providing a catalog of HI tidal remnants; directly determining 
the HI diameter function; including $\sim 2000$ continuum sources 
with fluxes sufficiently large to make useful measurements of HI optical depth, 
it will provide a low z link with DLA absorbers; dramatically
expand the targets for the study of environmental effects on the evolution
of galaxies, through the use of diagnostic tools such as gas deficiency;
significantly expand the redshift database for the dynamical study
of nearby groups and clusters; bringing to unprecedented flux limits the 
census of High Velocity Clouds and possibly clarifying their relationship to 
Local Group and other nearby galaxies.

As of early 2008, ALFALFA observations have been completed for approximately
40\% of the survey goal of 7000 square degrees. Data processing to "level I"
(see \cite{a1}) is complete for those data, data cubes and source extraction
is complete for approximately 25\% of the survey. Public access to cataloged
sources in digital form is available\footnote{http://arecibo.tc.cornell.edu/hiarchive/alfalfa}
through robust software tools.
Several preliminary results of ALFALFA are presented in these proceedings, with
particular emphasis on discoveries in the Virgo cluster region, which was given
priority in coverage in the early phases of the survey.
Possible developments with next generation HI surveys at Arecibo, as an SKA 
precursor, are discussed in the last session of these proceedings.

\begin{theacknowledgments}
This work has been supported by NSF grants AST--0307661,
AST--0435697, AST--0607007. The Arecibo 
Observatory is part of the National Astronomy and Ionosphere Center,
which is operated by Cornell University under
a cooperative agreement with the National Science Foundation.
\end{theacknowledgments}

\doingARLO[\bibliographystyle{aipproc}]
          {\ifthenelse{\equal{\AIPcitestyleselect}{num}}
             {\bibliographystyle{arlonum}}
             {\bibliographystyle{arlobib}}
          }
\bibliography{sample}

\end{document}